\documentstyle[12pt,aaspp4,flushrt,tighten]{article} 

\slugcomment{To appear in Astronomical Journal, April 1998 issue}

\newcommand{\HS}{\hbox{HS\thinspace 0551+7241}}
\newcommand{\HI}{\hbox{H{\sc i}}}

\newcommand{\HeI}{\hbox{He{\sc i}}}
\newcommand{\HeII}{\hbox{He{\sc ii}}}
\newcommand{\Hal}{\hbox{H$\alpha$}}
\newcommand{\Hbe}{\hbox{H$\beta$}}
\newcommand{\Hga}{\hbox{H$\gamma$}}
\newcommand{\Hde}{\hbox{H$\delta$}}
\newcommand{\kms}{\mbox{\rm km\thinspace s$^{-1}$}}

\begin{document}


\title{HS\thinspace0551+7241: A New Possible Magnetic Cataclysmic Variable\\
in the Hamburg/CfA Bright Quasar Survey}

\author{Danuta Dobrzycka\altaffilmark{1}, Adam Dobrzycki}

\affil{Harvard-Smithsonian Center for Astrophysics\\
60 Garden Street, Cambridge, MA 02138, USA\\
e-mail: [ddobrzycka,adobrzycki]@cfa.harvard.edu}

\and

\author{Dieter Engels and Hans-J\"urgen Hagen}
\affil{Hamburg University Observatory\\
Gojenbergsweg 112, D-21029 Hamburg, Germany\\
e-mail: [dengels,hhagen]@hs.uni-hamburg.de}

\altaffiltext{1}{also Copernicus Astronomical Center, Bartycka 18,
00-716 Warszawa, Poland}


\begin{abstract}

We present the analysis of the spectroscopic observations of a newly
discovered cataclysmic variable \HS. In 1995 the star's brightness
dropped by $\Delta B\sim 2.5$~mag and \HS\ entered a low state lasting
for $\sim$2~yr. The \Hal, \Hbe\ and \HeII(4686)\ emission line radial
velocity curves show evidence of short $\sim 50$~min fluctuations
superimposed on longer, $\sim$4~hour variations. We found similar
modulations in the line fluxes and the line equivalent widths. The
continuum magnitude light curve suggest that \HS\ may be an eclipsing
system. If the 4~hour period is related to the orbital period then the
mass function is $\sim0.138 M_\odot$. Relatively strong \HeII(4686)
emission and short fluctuation may suggest that the binary is an
intermediate polar. In that case the $\sim$50~min variations would
correspond to the Alfv\'en radius of $\sim3\times10^{10}$~cm.

\end{abstract}


\section{Introduction}

Cataclysmic variables (CVs) are short period ($P_{\rm orb} \sim
1-15$~hr) interacting binaries in which a Roche lobe filling K or M
dwarf transfers matter into a white dwarf companion. Some of these
binaries contain strongly magnetic ($B\sim 10^7$~G) white dwarfs and
the accreting matter is funneled radially onto the magnetic poles.
Standoff shocks are produced above the white dwarf surface which can
efficiently convert the infall energy to X-rays. In some of these CVs
--- polars or AM Her systems --- the magnetic field of the primary is
so strong that it generates the signature of magnetic accretion: large
linear and circular optical polarization, and the primary is secularly
phase-locked. In others --- intermediate polars, commonly called
DQ~Her stars --- the primaries have less powerful magnetic fields and
there is no detectable polarization at optical wavelengths. Magnetic
CVs show short timescale periodic variations (from $\sim 30$~s to
2~hr) in the optical light as well as in the X-ray emission. The
pulses are attributed to the rotation of the magnetic white dwarf (see
e.g.\ Patterson 1994; Dobrzycka, Kenyon \& Milone 1996). However, the
majority of CVs are weakly magnetized systems where the accreting
material has excess momentum which does not allow matter to fall
directly onto the white dwarf. This accreting matter loses angular
momentum gradually as it spirals inward, which leads to the formation
of an accretion disc. A boundary layer is formed between the rapidly
rotating disk and the white dwarf.

In this paper, we report the discovery of a new cataclysmic variable,
\HS\ ($\alpha_{1950}= 05^{\rm h} 51^{\rm m} 15.9^{\rm s}$,
$\delta_{1950}= 72^\circ 41^{'} 27^{''}$). This object was
originally selected as a $B=16.8$~mag quasar candidate during the
course of the Hamburg/CfA Bright Quasar Survey (Engels et al.\
1998). Quasar candidates selected from the objective prism spectra
taken with the Hamburg Schmidt telescope at Calar Alto, Spain are
observed spectroscopically with the 1.5-meter Tillinghast telescope at
Whipple Observatory on Mt.Hopkins, Arizona. As it is usually the case
with such a survey, several candidates turn out to be CVs. The
identification spectrum of \HS, obtained in 1995, revealed features
typical for a CV: \HI, \HeI, and \HeII\ emission lines on top of blue
continuum.

We present the first detailed study of the spectroscopic variations of
\HS. We describe our observations in Section~2. In Section~3 we
discuss the data and present arguments for \HS\ possibly being an
intermediate polar system. We summarize our results in Section~4.

\section{Observations}

In 1991 and 1994, when the Hamburg Schmidt plates were taken, \HS\ had
an estimated magnitude of $B=16.8$. The identification spectrum was
taken in 1995, and the star appeared to be fainter by as much as
$\sim$2.5~mag. We have been monitoring this object since then; the
star remained in the low state in 1996. During our observing run in
March 1997 we routinely took an exposure of \HS\ and found the star to
be bright again.

\subsection{Spectroscopic observations}

\begin{figure}
\begin{center}
\epsfysize=0.8\textheight
{\centering\leavevmode\epsfbox{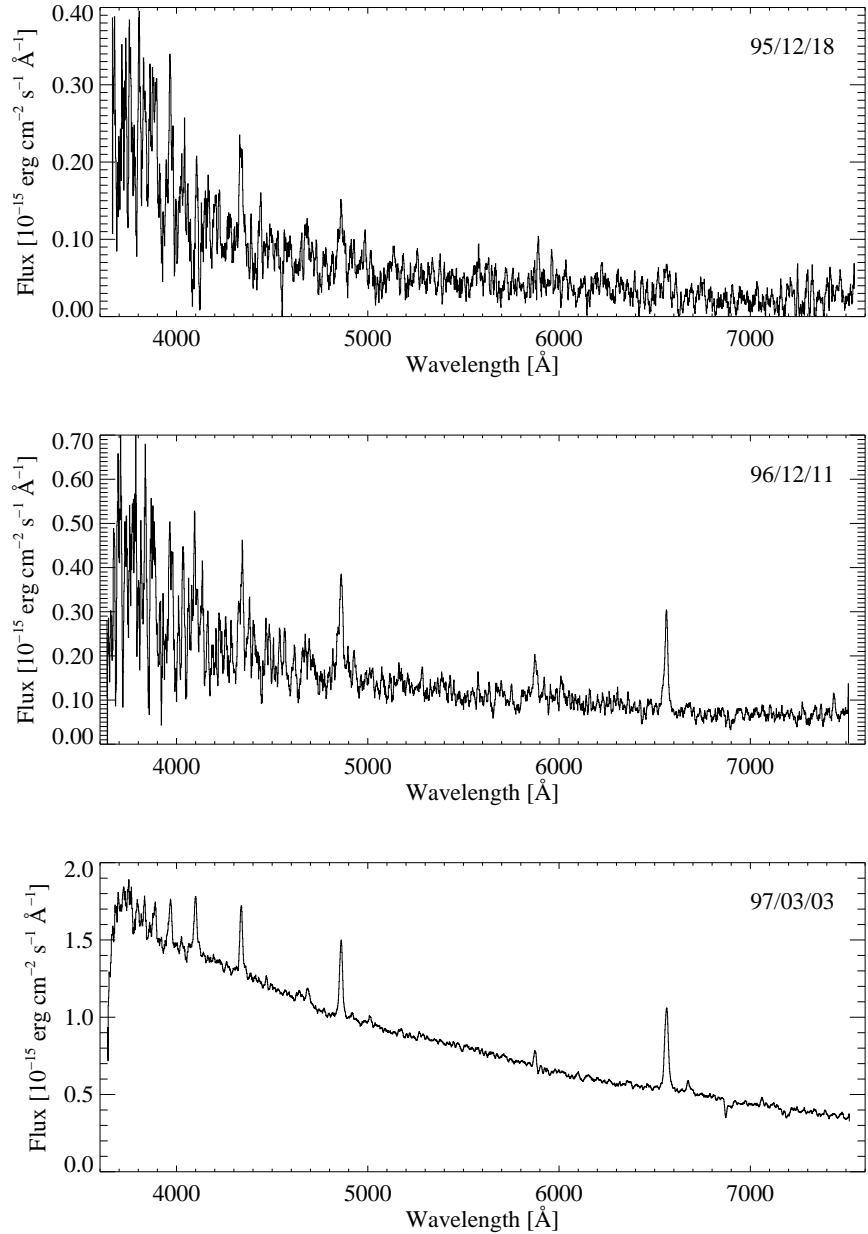}}
\caption{Typical spectra of \HS. The top and middle
panels show spectra obtained in the stars's low state, while the
bottom panel presents a spectrum acquired when the star returned to
its normal brightness.
\label{fig:typspe}}
\end{center}
\end{figure}

The spectra of \HS\ presented here were obtained with the FAST
spectrograph mounted on the 1.5-meter Tillinghast telescope at the
Fred Lawrence Whipple Observatory on Mt.~Hopkins, Arizona.  A thinned,
back-side illuminated $512\times 2688$ CCD served as the detector, 300
line/mm grating in first order and 3$^{\prime\prime}$ slit were used,
resulting in a dispersion of 100~\AA/mm and a resolution of 2.95\AA. A
typical spectral range was 3700--7400\AA. Each night bias, flatfield,
and He-Ne-Ar comparison lamp exposures were obtained and used for data
calibration. The 600~sec integration times were applied to secure good
S/N for the spectra. The weather was clear in all nights, with seeing
of $\sim 2-3^{\prime\prime}$. We acquired spectra of \HS\ with the
same instrument on several occasions in 1995 and
1996. Table~1 shows the journal of observations. All
the spectroscopic data were reduced to the Hayes \& Latham (1975) flux
scale using standard packages within IRAF. Photometric calibration of
data obtained in 1995 has an uncertainty of $\pm 0.3-0.5$~mag, in 1996
$\pm 0.1-0.3$~mag and in 1997, $\pm0.05-0.07$~mag. See
Figure~\ref{fig:typspe} for examples of \HS\ spectra.

\begin{figure}
\begin{center}
\epsfxsize=\textwidth
{\centering\leavevmode\epsfbox{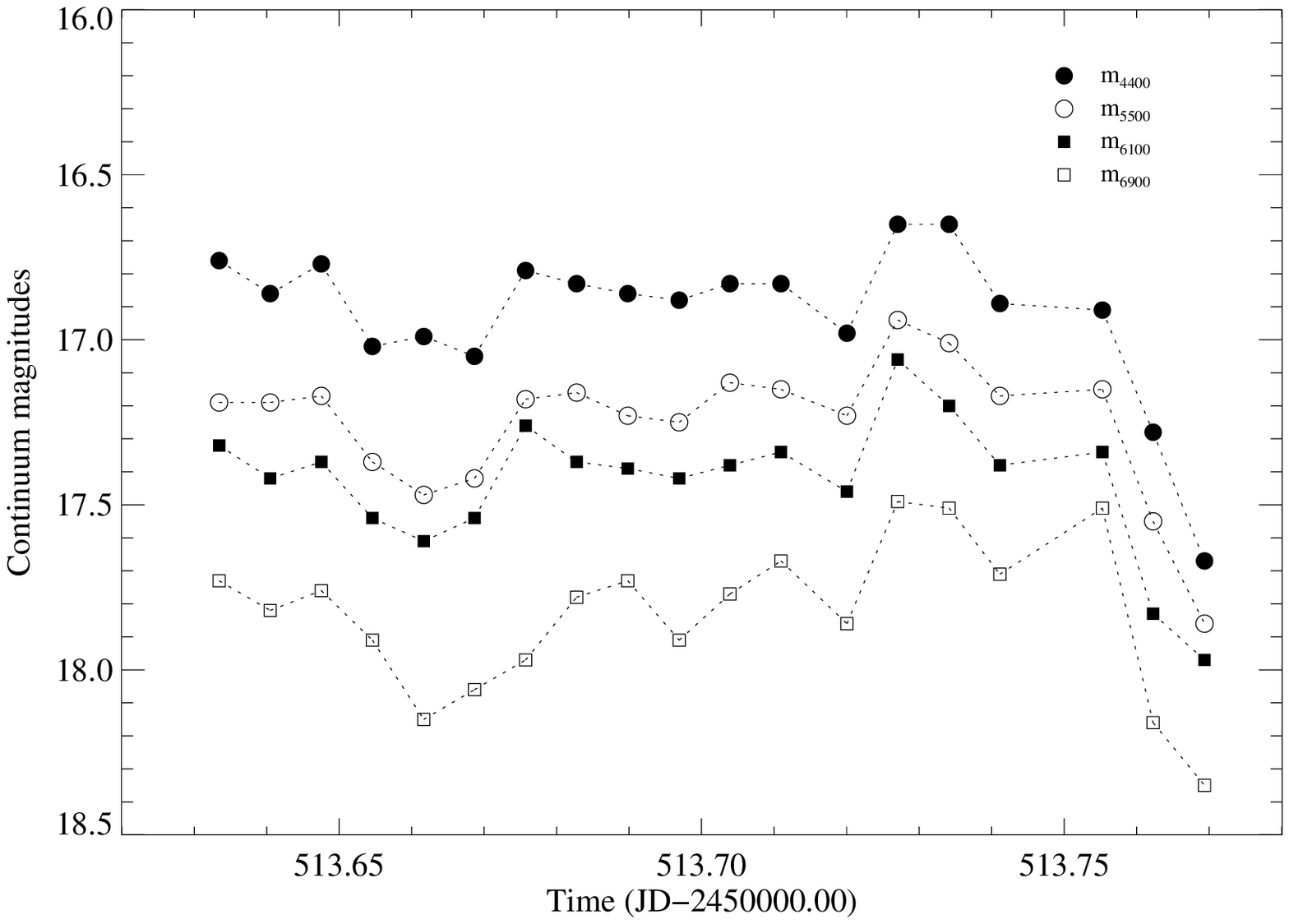}}
\caption{Continuum magnitudes of \HS\ obtained on
March 05, 1997. At the end of our observing sequence all the continuum
magnitudes decreased by $\sim 1$~mag, which may indicate beginning of
the primary eclipse.
\label{fig:contmags}}
\end{center}
\end{figure}

To search for the continuum variations we measured average fluxes in
20\AA\ bins free from emission lines and derived narrow band continuum
magnitudes, $m_\lambda = 2.5~ \log F_\lambda - 21.1$, where
$F_\lambda$ is in erg~cm$^{-2}$~s$^{-1}$\AA$^{-1}$.
Figure~\ref{fig:contmags} shows the continuum magnitudes derived from
the observations on March 05, 1997.

In addition to the narrow band continuum magnitudes, we also estimated
average broad band B- and V-like magnitudes. We combined all the
spectra obtained during each night and calculated $m_B = -2.5 \log F_B
- 20.45$, and $m_V = -2.5 \log F_V - 21.1$, where $F_B$ and $F_V$ are
the average fluxes in the spectral regions corresponding to the width
of the B and V bands (Allen 1973). The results are listed in
Tables~1 and 2.

\begin{figure}
\begin{center}
\epsfysize=0.8\textheight
{\centering\leavevmode\epsfbox{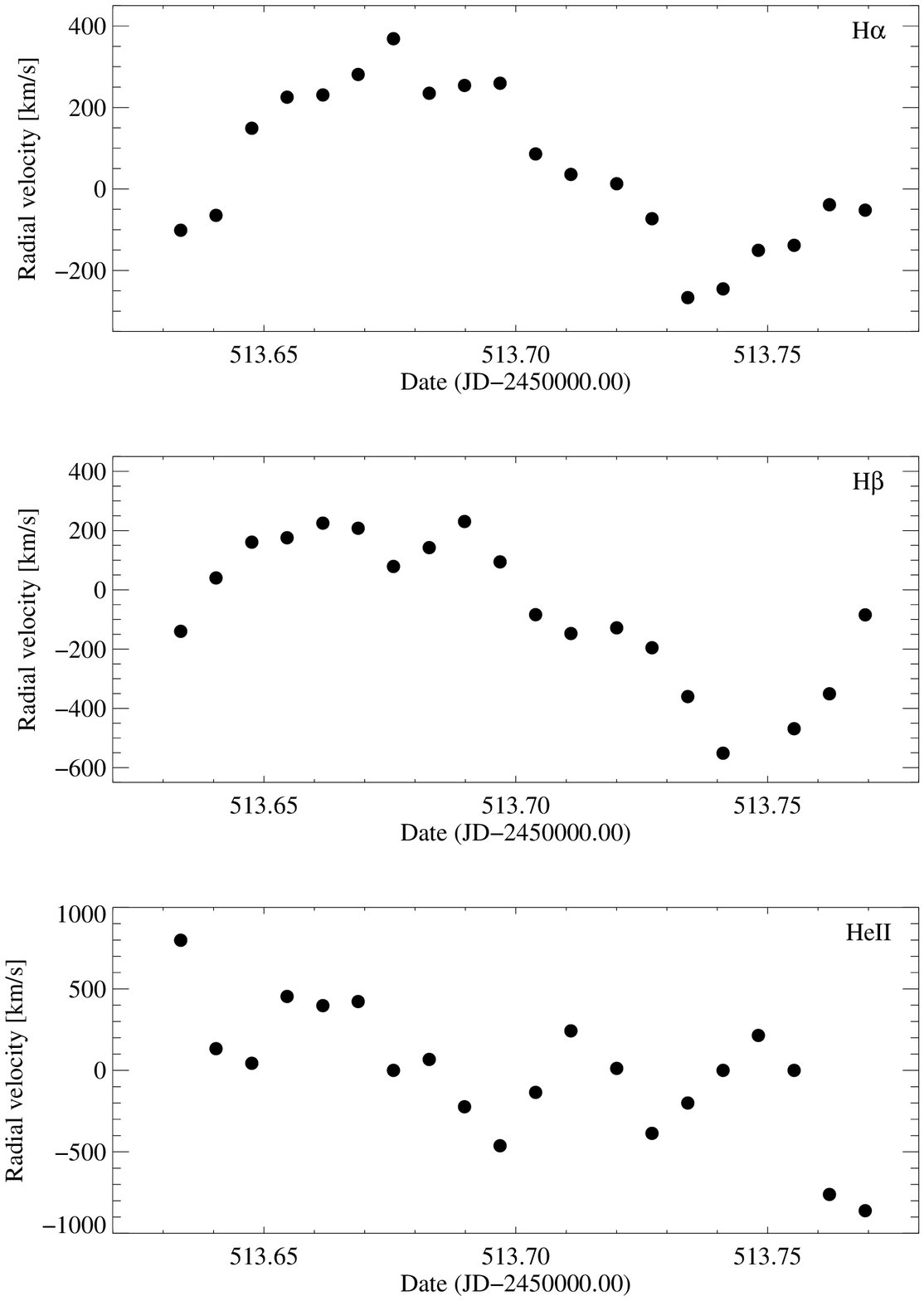}}
\caption{Radial velocity data for \HS\ obtained on
March 05, 1997. Top panel shows the radial velocities derived from the
\Hal\ emission line, middle --- \Hbe\, bottom --- \HeII(4686).
\label{fig:radvel}}
\end{center}
\end{figure}

\begin{figure}
\begin{center}
\epsfysize=0.8\textheight
{\centering\leavevmode\epsfbox{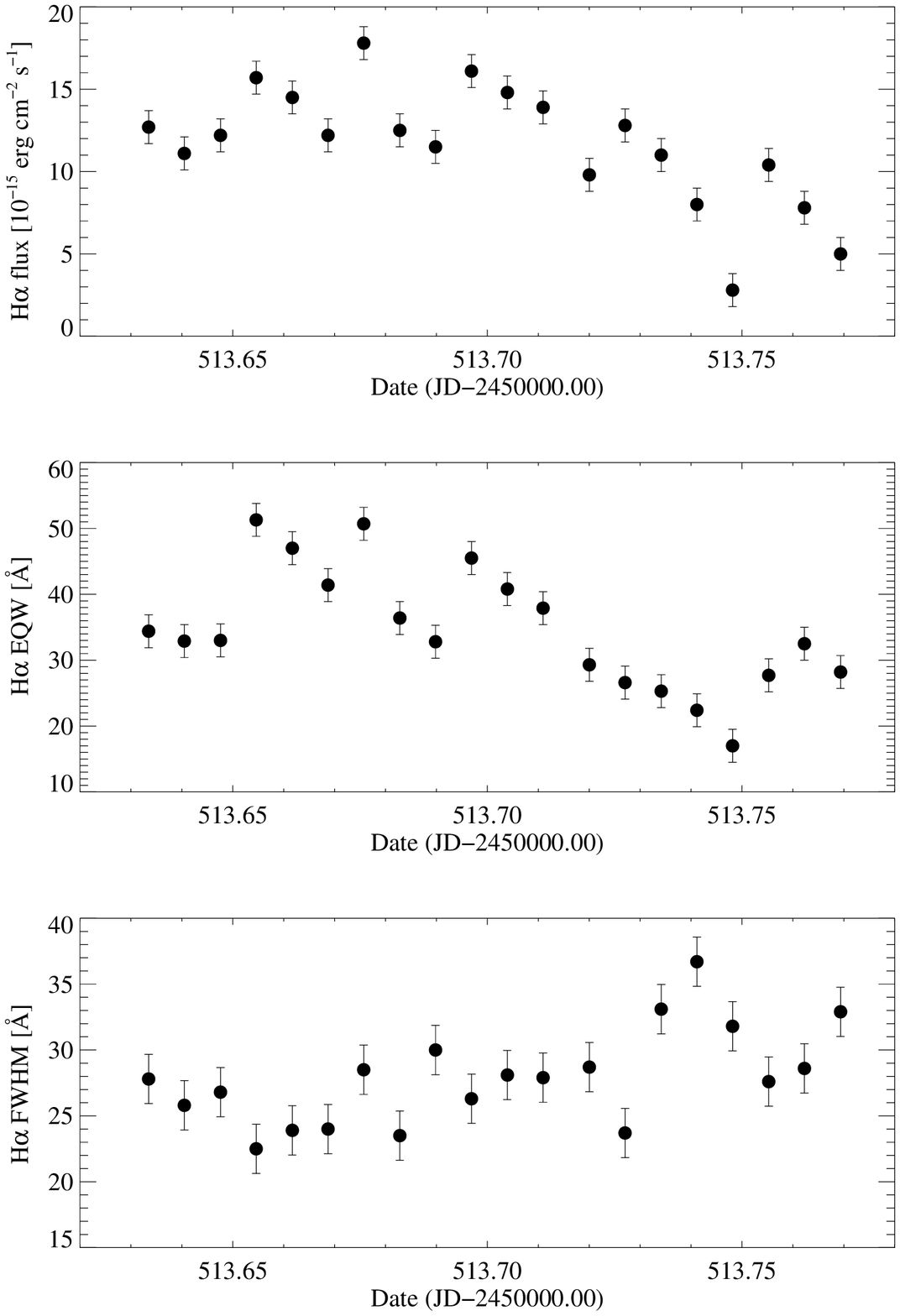}}
\caption{Variations of the \Hal\ emission line
during our observations on March 05, 1997. The top panel shows
variations of the line flux, the middle and lower panels plot
variations of the line's equivalent width and
full-width-at-half-maximum, respectively.
\label{fig:varemis}}
\end{center}
\end{figure}

We measured emission line fluxes, equivalent widths (EQW) and FWHM
values by fitting Gaussian profiles. We estimate the uncertainties of
the line fluxes to be between $\sim15-25$\% for data obtained in
1996,1997 and $\sim20-35$\% in 1995, depending on the line strength
and fluctuations in the local continuum level.  The uncertainties in
the estimates of EQW and the FWHM of the emission lines are $\sim
15-20$\% and these quantities were derived for the 1996 and 1997 data
only. We also measured radial velocities for the \Hal, \Hbe\ and
\HeII(4686) lines assuming that the center of the fitted Gaussian is
the center of the line. We found their typical uncertainties to be
$\sim 20-30$~\kms. The \HeII(4686) line lies close to the C{\sc
iii}/N{\sc iii} blend at 4650\AA. In most cases we were able to
resolve these lines and to fit separate Gaussians to each of them, but
that could, however, cause larger uncertainties of the measurements of
the \HeII(4686) line. We estimate the flux, EQW and FWHM uncertainties
for this line to be $20-25$\% and the radial velocity uncertainties
$\sim 30-40$~\kms. Table~3 and
Figures~\ref{fig:radvel} and \ref{fig:varemis} show derived properties
of the emission lines. The data were not corrected for the
interstellar reddening.

Figure~\ref{fig:typspe} shows examples of typical spectra of \HS: two
obtained in 1995 and 1996, in the star's low state, and one obtained
in 1997, when the star became brighter again. It can be clearly seen
that in the low state the continuum was much fainter, by $\Delta m_B
\approx 2.5$~mag in 1995, and by $\Delta m_B \approx 1.8$~mag in
1996. Decreasing amplitude may suggests that in late 1996 the star was
already rising from its low state. All spectra display blue continuum,
strong \HI\ Balmer lines, \HeI\ at 5876\AA, 6678\AA, 4471\AA, and
4922\AA, \HeII\ at 4686\AA\ and the C{\sc III}/N{\sc III} blend at
4650\AA. In the low state the emission lines are generally weaker than
in 1997 and \Hga\ appears to be the strongest emission feature. In
most of the CVs the optical spectrum of the secondary appears more
clearly during the low state. Since the low state spectra of \HS \
have relatively low signal-to-noise we are not able to look for
absorption features characteristic for the secondary.

\subsection{Previous photometric observations of \HS.}

We attempted to obtain photometry of \HS\ in October and December
1996. On October 21 the McGraw-Hill 1.3-meter telescope at the MDM
Observatory with the CCD camera was used; remaining observations
(October 28-30 and December 17-18, 1996) were performed with the the
FLWO 1.2-meter telescope with the CCD camera. The October observations
were obtained by K.~Stanek. We used standard UBVR filters with
response curves close to those listed in Bessell (1990) and
auto-guided every observation. On both occasions the object appeared
to be much fainter than B~$\sim$~16.8~mag, which was its brightness
when it was originally discovered on the Schmidt plates. We were able
to place only upper limit of B~$\ga$~18.7~mag. We estimated the upper
limit for the brightness of \HS, judging from the detection of other
faint stars in the field of view, which had B magnitudes ranging from
17 to 18.7~mag (see below).

Since \HS\ seems to undergo low and high states it is interesting to
investigate its light variations from archival observations. We
collected the available estimates of the star brightness; we list
them, together with our data, in Table~2.

The object is relatively faint, and so we were able to detect it only
on the plates from the MC series of the sky patrol plates in the
Harvard plate archive. These plates have limiting magnitude of
17~mag. The plates were taken in a photographic bandpass; we related
the $m_{\rm pg}$ magnitudes from the plates to the $B$ magnitude using
the formula from Allen (1973), $m_{\rm pg}=B-0.11$~mag. On each plate
we looked for \HS\ and for several neighboring stars, which had
brightness ranging from $m_{\rm pg}=16.4$ to 18.6~mag. Their
magnitudes were derived from the Digitized Sky Survey and have
uncertainties of $\sim 0.1$~mag. As they cover relatively wide range
of magnitudes, we were able to put some limits on the brightness of
\HS. All the examined plates showed the CV to be brighter than $m_{\rm
pg}\sim 17$~mag ($B\sim 16.9$~mag) indicating that it was in the high
state. The results of our search are summarized in
Table~2.

\section{Discussion}

The random historical observations of \HS\ from the first decades of
the century showed that its brightness remained comparable to that
seen on the Schmidt plates --- $B=16.8$~mag. Thus, it can be assumed
that this is the {\em normal} brightness of the star, i.e.\ the
brightness at which the star remains for most of the time. We were
fortunate to observe \HS\ fading down by $\Delta B\sim 2-3$~mag to its
low state between January 1994 and December 1995. In late 1996 the
star already seemed to be rising from the minimum. During our
observations in March 1997 the \HS\ brightness was fluctuating from
day to day with amplitude in $B$ and $V$ of $\sim0.5$~mag
(Table~1). Such amplitude modulation in brightness is
typical for most CVs (Warner 1995).

The interesting feature in the low state spectra is the unusual
relative strength of \HI\ Balmer lines. While the spectrum from 1995
has relatively low S/N ratio, it can be clearly seen that \Hga\ is the
strongest emission feature. Reddening towards \HS\ is not yet known in
detail, however we estimate it to be $E_{B-V} \le 0.12$. This estimate
is based on the total extinction through the Galaxy from the galactic
reddening map of Burstein \& Heiles (1982). For this reddening and the
standard reddening law (Savage \& Mathis 1979), the ratio of \Hbe\ and
\Hal\ lines is $I(\Hbe)/ I(\Hal) \sim 1.5-1.3$ in 1995, 1996 and $\sim
1.0-0.7$ in 1997, while for the optically thin gas we would rather
expect it to be around 0.3.

Flat Balmer decrement, implying that the Balmer emission lines are
optically thick, has been observed in most CVs, including intermediate
polars (Horne \& Marsh 1986, Warner 1995). We see it in the 1997
data. However, the inverted decrement, possibly present in the low
state spectra, is typical for AM Her type of magnetic CVs and is also
occasionally observed in intermediate polars (e.g.\ EX~Hya, Williams
1983; TW Pic, Mouchet et al.\ 1991). To achieve inverted flux ratio,
the collisional effects between $n=2$ and $n=3$ levels must be
important, implying the electron density of $N_e \ge
10^{12}$~cm$^{-3}$ in the line emitting regions (Stockman et al.\
1977; Liebert \& Stockman 1985).

In the spectrum from 1995 there is also an evidence of strong
\HeII(4686). Its strength seems to decrease as the star gets brighter
and $I(\HeII4686)/I(\Hbe)$ approaches $\sim$0.5 in 1996 and 1997.
Strong \HeII(4686) line has often been used as an indicator of the
magnetic field. For the strong AM~Her systems this is a confirmed
attribute (Liebert \& Stockman 1985, Warner 1995).  However, Szkody
(1995) pointed out that previous analyses showed that the \HeII\
strength in confirmed DQ~Hers shows a large range, from the flux much
greater than \Hbe\ at one extreme, to zero at the other. Silber (1986)
found a good way to separate magnetic from non-magnetic CVs by
plotting $I(\HeII4686)/I(\Hbe)$ ratios versus the EQW of \Hbe. In this
plot, the systems with the ratio $\ge 0.4$ and the EWQ(\Hbe)~$\geq$~20
are magnetic. Since this method applies to magnetic objects in the
high state we can only analyze the 1997 spectra. Using $E_{B-V}\leq
0.12$, we obtain $I(\HeII4686)/I(\Hbe) \approx 0.7 - 0.3$ and
EQW(\Hbe)$\approx 25-10$~\AA. It is worth noting that the
\HeII(4686)/\Hbe\ ratio is in practise not sensitive to reddening and
thus it is not affected by the uncertain value of $E_{B-V}$. Clearly,
\HS\ sometimes does and sometime does not fulfill the Silber's
criterion for a magnetic CV.

We observed \HS\ undergoing a low state. The optical brightness of the
star decreased by $\Delta B \sim 2-3$~mag during the interval of $\sim
2$ years. Similar light variations are typical for magnetic CVs, both
polars and intermediate polars (Garnavich \& Szkody 1988). Long term
light curves of, e.g., AM~Her, AN~UMa, and KO~Vel (Feigelson, Dexter
\& Liller 1978; Meinunger 1976; Cropper 1986; Mukai \& Corbet 1987)
show that these systems usually stay at the maximum brightness, but
occasionally experience drop to a lower brightness. The typical
amplitudes are $\Delta B\sim 1.5-5$~mag and the stars remain in the
low states for few years. The significant long term reduction in
optical luminosity is naturally interpreted as a lowering of the
average mass transfer rate. However, other CVs that undergo
similar variations are nova-like variables, of VY~Scl type (Garnavich
\& Szkody 1988; Warner 1995). With our observations we are not able to
unambiguously distinguish if the low states of \HS\ point to the
magnetic or the VY~Scl type variations.

The \Hal\ and \Hbe\ radial velocity curves of \HS\
(Figure~\ref{fig:radvel}) appear to consist of small, short
fluctuations superimposed on larger sinusoidal variations. The short
fluctuations seem to dominate the \HeII(4686) radial velocity
curve. Also, long and short fluctuations seem to be present in the
hydrogen emission line fluxes, EQW and FWHM
(Figure~\ref{fig:varemis}). The rapid variations may also be seen in
the \HeII(4686) line properties (Table~3). However, the uncertainties
are quite large and the variations are not obvious.

We analyzed the whole sample of radial velocities as well as the
emission line properties using Monet's (1979) Fourier transform
algorithm (Kenyon \& Garcia 1986) and the Lehmann-Filhes method (e.g.\
Petrie 1962). The best large-amplitude modulations that we found had a
timescale of $\sim 4$~h.  The period of $4.01\pm0.4$~h was present in
both, \Hal\ and \Hbe, radial velocity curves, as well as in the \Hbe\
intensities. The \Hal\ intensities had shorter period of
$3.79\pm0.41$~h, while EQW of \Hal\ and \Hbe\ had a longer period of
$4.20\pm0.65$~h. However, they all agree with the 4~h fluctuations.
We have to mention that the periods we found are close to the
characteristic length of our data sets, and as such have to be taken
with caution. We also do not have enough data to fold them with this
period. More observations are needed to confirm the longer
modulations. The periodogram analysis of the \HeII(4686) radial
velocity curve showed evidence of the period of $50\pm5$~min and the
longer period of $4.48\pm0.81$~h supporting the 4~h modulations found
earlier. Similarly, $\sim 50$~min fluctuations were found in the \Hbe\
flux. We were not able to look for shorter variations as the time
resolution of our data only allows us to see variations with
characteristic timescales of tens of minutes. No other significant
periods were found in the data. The continuum magnitude light curves
do not show any obvious periodical modulations.

Superposition of longer and shorter periods in the \HS\ data, where
the longer modulations cannot be described in detail, does not allow
us to subtract it and search more carefully for the shorter period in
residuals. Future observations of \HS\ may provide us with better data
for more detailed periodogram analysis.

Similar long and short variations are characteristic for intermediate
polars. High quality, timed-resolved spectroscopic studies of e.g.\
EX~Hya (Hellier et al.\ 1987), FO~Aqr (Hellier, Mason \& Cropper 1990;
Martell \& Keitchuck 1991), AO~Psc (Hellier, Cropper \& Mason 1991),
showed complex structure of the emission lines. The radial velocities
of the Balmer lines can be dominated by the S-wave component following
the orbital period, while the \HeII(4686) line radial velocities show
variations with a rotational period of a magnetic white dwarf or a
beat period. The short modulations also appear in the fluxes and EQW
of the emission lines.

Assuming that the detected $\sim$4~hr period corresponds to the
orbital period, and the 50~min pulses to the rotational period of
the primary we can put some constraints on the system parameters.
Fitting circular orbit to the radial velocities yields large scatter
in the systemic velocities, $\gamma$. Semi-amplitudes range from
$K_1\sim173$~\kms\ for \Hal\ up to $K_1\sim 220$~\kms\ for \Hbe\ and
\HeII(4686).

The radial velocities of the Balmer lines have been commonly used for
the spectroscopic orbit determination for many CVs (see e.g.\
Dobrzycka \& Howell 1992). However, in non-magnetic CVs the \HI\
emission usually originates in the outer parts of the accretion disk
and do not therefore trace accurately the orbital motion of the white
dwarf. In magnetic CVs the Balmer lines usually originate in gas
falling down towards the accretion zone (polars) or in the hot spot
(intermediate polars). Figure~\ref{fig:contmags} shows the continuum
magnitude light curves obtained for \HS\ on March 05, 1997. At the end
of our observing sequence the continuum brightness decreased at all
wavelengths and it appears as if the system was about to enter an
eclipse. We estimated following drops of brightness: $\Delta
m_{4400}\approx 0.86$~mag, $\Delta m_{5550}\approx 0.70$~mag, $\Delta
m_{6100}\approx 0.63$~mag, and $\Delta m_{6900}\approx 0.58$~mag. This
suggests that continuum was getting redder what could indicate
possible eclipse of the white dwarf by the secondary. The H~I emission
lines weaken, but do not disappear completely, while at the same time
the He~II line becomes hardly visible. If what we observe is a real
primary eclipse then the radial velocities of the Balmer lines are out
of phase, which suggests that they follow the orbital motion of the
secondary, and not of the white dwarf. The radial velocity curve of
\HeII(4686) is dominated by the short fluctuation and analysis of the
orbital variations is very unreliable. Another conclusion that can be
drawn from the assumption that \HS\ is an eclipsing binary is that the
inclination angle of the system's orbit, $i$, is close to $90^\circ$.

Since all derived semi-amplitudes agree within errors we adopt average
value of $K_1=200$~\kms\ in further considerations. In view of above
discussion we want to emphasize that determination of system
parameters for \HS\ is uncertain and give us only rough
estimates. Applying adopted values, $P=4$~h, $K_1=200$~\kms\ and $i
\approx 90^\circ$, to the mass function equation, $f(M_2)=M_2
[q/(q+1)]^2 \sin^3i$ we find $f(M_2)\approx0.138 M_\odot$. For an
average mass of the white dwarfs in magnetic CVs with $P_{\rm
orb}\ge2.4$~h, $M_1\approx 0.75~M_\odot$ (Webbink 1990) we obtain mass
ratio $q=M_2/M_1 \sim 0.7$ and thus $M_2\approx 0.5~M_\odot$. This is
in a good agreement with the $M_2\approx 0.4 M_\odot$ calculated from
the mean empirical mass-period relationship $M_2 = 0.065 P_{\rm
orb}(h)^{1.25}$ (Warner 1995).

In the intermediate polars the accretion disk extends outward from the
magnetosphere of the white dwarf. The magnetosphere is defined by the
{\em Alfv\'en radius}, at which the pressure exerted by the infalling
material is balanced by the magnetic pressure. Assuming spherical,
free-fall accretion, the Alfv\'en radius can be written as $R_A =
5.1\times10^8 {\dot M}_{16}^{-2/7} M_1^{-1/7} \mu_{30}^{4/7}$~cm,
where ${\dot M}_{16}$ is the mass accretion rate in units of
$10^{16}$~g~s$^{-1}$, $M_1$ is expressed in $M_\odot$, and $\mu_{30}$
is the magnetic moment in units of $10^{30}$~G~cm$^3$ (Frank, King \&
Raine 1992). The magnetic moment is specified by the white dwarf
surface field strength, $B$, and is expressed with $\mu = B
R_1^3$. For typical ${\dot M} \approx 10^{16}-10^{17}$~g~s$^{-1}$ and
$B\sim 10^7$~G we obtain $R_A\approx 3\times 10^{10}$~cm if we use
$M_1=0.75~M_\odot$ and the white dwarf radius of $R_1\approx
5\times10^8$~cm. As the accreting gas finds its way inside the
Alfv\'en radius, it produces shock heated region on the inner edge of
the disk. After that, the matter flows along the magnetic field lines
towards the magnetic poles of the white dwarf. If the base of the
accretion stream on the Alfv\'en radius remains fixed with respect to
the disk then the heated region will rotate with the Keplerian
velocity of the inner edge of the disk. The detected 50~min period
corresponds to the radius of a Keplerian orbit $R\approx
2.9\times10^{10}$~cm, which agrees very well with estimated Alfv\'en
radius. We can compare it with the semiamplitude of the white dwarf's
orbit $A_1 \sin i= (2\pi)^{-1} P_{\rm orb} K_1\approx
5\times10^{10}$~cm; the Alfv\'en radius lies well within it.

\section{Conclusions}

We observed a new cataclysmic binary \HS. Our analysis of
spectroscopic and historic photometric data showed that this object
may be a candidate for the intermediate polar.

We observed \HS\ undergoing a low state lasting $\sim 2$~yr. During
the minimum of its activity the stellar brightness dropped by $\Delta
B\sim 2-3$~mag; such low states are common among AM~Her and DQ~Her
systems. The periodogram analysis of the \Hal, \Hbe\ and \HeII(4686)
radial velocities revealed that the hydrogen lines are likely
dominated by the S-wave component, following the orbital variations
with $P_{\rm orb}\sim 4$~hr, while \HeII(4686) shows rapid
fluctuations with $P\approx 50$~min. Similar long- and short-term
variations were present in the emission line fluxes and EQW. Adopting
4~hr modulation as the \HS\ orbital period we obtain the mass function
$f(M_2)\approx 0.138~M_\odot$, and, after some reasonable assumptions,
we put some constraints on the system parameters: $M_1 =0.75~M_\odot$,
$q=M_2/M_1 \sim 0.7$, $i\sim 90^\circ$. The 50~min period very likely
corresponds to the white dwarf's Alfv\'en radius $R_A\sim
3\times10^8$~m.

Polars and intermediate polars are known to be relatively bright X-ray
sources, although there are several objects that have not been
detected in X-rays (Patterson 1994). To our knowledge, no X-ray
sources were detected on the position of \HS, including the Rosat All
Sky Survey (Greiner \& Voges 1997).

HS\thinspace0551+7241 appears to be a very interesting object. More
UV/optical/IR and polarimetric observations are needed to derive
accurately the system parameters, characterize properties of the white
dwarf and the red dwarf companion, and to probe the magnetic nature of
the system. Future pointed X-ray observations could provide us with
additional informations about the short signal modulations and put
more light on the physical properties of the primary.

\acknowledgements

We thank the staff of the FLWO 1.5-m telescope for assistance with
obtaining the observations, K.~Stanek for acquiring optical photometry
in October 1996, and S.~Kenyon for providing us with the code for the
periodogram analysis. We also thank M.~Hazen, M.~Livio, P.~Garnavich,
E.~Schlegel, and the anonymous referee for advice and helpful
comments.  This research has made use of the Simbad database, operated
at CDS, Strasbourg, France, and of the Digitized Sky Survey, produced
at the Space Telescope Science Institute under U.S. Government grant
NAG W-2166. AD acknowledges support from NASA Contract No.\ NAS8-39073
(ASC). The Hamburg Quasar Survey is supported by the DFG through
grants Re~353/11 and Re~353/22.



\newpage

\footnotesize

\begin{center}
\begin{deluxetable}{cccccc}
\tablenum{1}
\tablecaption{Journal of observations.}
\tablewidth{0pt}
\tablehead{
\colhead{Date} & \colhead{JD} & \colhead{Exp.Time [s]} & 
\colhead{Duration [h]} & \colhead{Avg.\ $m_B$ [mag]} &
\colhead{Avg.\ $m_V$ [mag]}
}
\startdata
1995/12/18 & 2450070.81163 & 400 & 0.11 & 19.3 & 19.5 \nl
1996/12/11 & 2450429.75818 & 600 & 0.17 & 18.6 & 18.7 \nl
1997/03/02 & 2450510.69231 & 480 & 4.66 & 17.0 & 16.9 \nl
1997/03/03 & 2450511.65205 & 600 & 2.19 & 16.7 & 16.7 \nl
1997/03/05 & 2450513.70140 & 600 & 3.30 & 17.4 & 17.2 \nl
\enddata
\end{deluxetable}
\end{center}

\begin{center}
\begin{deluxetable}{ccl}
\tablenum{2}
\tablecaption{Previous observations of \HS.}
\tablewidth{0pt}
\tablehead{
\colhead{Date} & \colhead{Magnitude} & \colhead{Source}
}
\startdata
1916/10/01  & $B\sim 17$& Harvard plate MC11258 \nl
1935/01/02  & $B\sim 17$  & Harvard Plate  MC27582 \nl
1935/01/05  & $B\sim 17$  & Harvard Plate  MC27594 \nl
1935/02/05  & $B< 17$  & Harvard Plate  MC27640 \nl
1935/12/21  & $B\sim 17$  & Harvard Plate  MC28038 \nl
1991/04/15  & $B=16.8$  & Hamburg Plate H1995 \nl
1994/01/13  & $B=16.8$  & Hamburg Plate H2621 \nl
1995/12/18  & $m_B=19.3, m_V=19.5$ & FLWO 1.5~m telescope, FAST \nl
1996/10/28  & $B>18.7$ & MDM 1.3~m telescope, CCD camera \nl
1996/10/29  & $B>18.7$ & MDM 1.3~m telescope, CCD camera \nl
1996/10/30  & $B>18.7$ & MDM 1.3~m telescope, CCD camera \nl
1996/12/11  & $m_B=18.6,m_V=18.7$ & FLWO 1.5~m telescope, FAST \nl
1996/12/17  & $B\ga18.7$ & FLWO 1.2~m telescope, CCD camera\nl
1996/12/18  & $B\ga18.7$ & FLWO 1.2~m telescope, CCD camera \nl
\enddata
\end{deluxetable}
\end{center}

\scriptsize

\begin{center}
\begin{deluxetable}{ccccccc}
\tablenum{3}
\tablecaption{Optical emission line fluxes in units of $10^{-15}$
erg~cm$^{-2}$~s$^{-1}$.}
\tablewidth{0pt}
\tablehead{
\colhead{JD (2450000.0+)} & \colhead{f(\Hal)} & \colhead{f(\Hbe)} &
\colhead{f(\Hga)} & \colhead{f(\Hde)} & \colhead{f(\HeI5876)} & 
\colhead{f(\HeII4686)}
}
\startdata
070.81163& 2.1 & 2.7 & 3.2 & 2.6 & 0.8 & 2.0 \nl 
429.75818& 4.4 & 5.1 & 3.5 & \nodata & 1.7 & 2.4 \nl 
510.59532& 5.5 & \nodata & \nodata & \nodata & \nodata & \nodata \nl 
510.60044& 5.8 & 7.9 & 5.5 & \nodata & \nodata & \nodata \nl
510.61924&10.0 & 7.5 & 5.9 & 4.8 & 3.6 & 1.7 \nl 
510.64208&10.7 & 9.6 & 8.4 & 7.2 & 1.7 & 1.7 \nl 
510.66504&14.7 & 7.1 & 5.2 & 3.9 & 2.6 & 2.4 \nl 
510.68736&12.8 & 7.7 & 5.7 & \nodata & 1.2 & 3.3 \nl 
510.71493&10.5 & 6.6 & 6.7 & \nodata & 1.6 & 2.4 \nl 
510.74112&10.8 & 6.4 & \nodata & 3.9 & 1.3 & 3.1 \nl 
510.76408& 8.4 & 7.2 & 7.1 & 3.9 & \nodata & 1.6 \nl 
510.78929&12.7 &10.5 & 9.3 &12.5 & 2.8 & \nodata \nl 
511.60733&16.6 &11.5 & 8.7 & 6.8 & \nodata & 3.7 \nl
511.61478&13.0 & 9.5 & 4.7 & 6.0 & 1.0 & 2.4 \nl 
511.62222&11.3 & 6.8 & 7.3 & 8.5 & 4.4 & 4.9 \nl 
511.62973&11.9 & 9.4 & 6.5 & 5.6 & 5.4 & 4.6 \nl
511.63720&10.9 &10.2 & 8.4 & 6.8 & 1.6 & 2.0 \nl
511.64463&16.9 &20.4 &17.2 &17.2 & 5.2 & 6.0 \nl 
511.65206&14.8 &13.0 &12.7 &10.0 & 4.7 & 3.7 \nl 
511.65948&12.0 &11.2 & 9.2 & 6.8 & 3.8 & 5.7 \nl 
511.66690&13.1 & 9.3 & 7.4 & 5.2 & 3.0 & 3.9 \nl 
511.67434&14.2 & 8.0 & 5.9 & 3.2 & 3.7 & \nodata \nl 
511.68179&12.6 & 6.0 & 3.0 & 4.1 & 0.6 & \nodata \nl 
511.68925&12.0 & 9.1 &10.0 & 9.4 & \nodata & 2.7 \nl 
511.69676&12.7 & 8.2 & 4.6 & 5.8 & 1.7 & \nodata \nl 
513.63348&12.7 & 9.4 &10.4 & 7.6 & 3.1 & 1.9 \nl 
513.64052&11.1 & 8.0 & 6.9 & 5.3 & 1.5 & 3.6 \nl
513.64757&12.2 & 8.9 & 9.5 & 4.3 & 2.9 & 2.9 \nl 
513.65462&15.7 &13.0 &10.5 &15.1 & 2.8 & 2.8 \nl 
513.66166&14.5 & 9.8 & 7.4 & 5.4 & 1.9 & 3.7 \nl 
513.66870&12.2 & 9.3 & 3.2 & 4.1 & 3.0 & 2.4 \nl 
513.67574&17.8 &11.9 & \nodata &10.0 & \nodata & \nodata \nl 
513.68279&12.5 & 6.7 & 6.3 & \nodata & \nodata & 3.3 \nl 
513.68983&11.5 & 9.9 & 8.7 & 3.7 & 2.0 & 2.8 \nl 
513.69687&16.1 &11.0 & 8.1 & 7.0 & 2.4 & 0.9 \nl 
513.70392&14.8 &10.6 & 6.5 & 4.6 & 1.4 & 2.2 \nl 
513.71096&13.8 & 7.0 & 6.8 & 4.9 & 1.7 & 1.2 \nl
513.72002& 9.8 & 6.4 & 5.8 & \nodata & 1.1 & 3.5 \nl 
513.72707&12.8 & 9.4 & 7.2 & \nodata & 2.6 & 2.1 \nl 
513.73411&11.0 & 3.9 & 3.7 & \nodata & 0.8 & 1.6 \nl
513.74116& 8.0 & 4.5 & 3.8 & \nodata & 0.6 & \nodata \nl 
513.74819& 2.8 & 1.8 & \nodata & \nodata & \nodata & 0.5 \nl 
513.75524&10.4 & 5.1 & 5.5 & \nodata & 3.0 & \nodata \nl
513.76228& 7.8 & 7.8 & 3.3 & \nodata & 1.8 & 2.9 \nl 
513.76932& 5.0 & 5.3 & 5.0 & $>$3.0 & 1.7 & 0.9 \nl
\enddata
\end{deluxetable}
\end{center}


\begin{references}

\reference{} Allen, C. W. 1973, Astrophysical Quantities, 3rd ed.\
(Athlone, London)

\reference{} Bessell, M. S. 1990, \pasp, 102, 1181

\reference{} Burstein, D., \& Heiles, C. 1982, \aj, 87, 1165

\reference{} Cropper, M. S. 1986, \mnras, 222, 225

\reference{} Dobrzycka, D., \& Howell, S. B. 1992, \apj, 614

\reference{} Dobrzycka, D., Kenyon, S. J., \& Milone, A. A. E. 1996,
\aj, 111, 414

\reference{} Engels, D., Dobrzycki, A., Hagen, H.-J., Elvis, M.,
Huchra, J., \& Reimers, D. 1998, in preparation

\reference{} Feigelson, E., Dexter, L., \& Liller, W. 1978,
\apj, 222, 263

\reference{} Frank, J., King, A. R., \& Raine, D. J. 1992, Accretion
Power in Astrophysics (Cambridge University Press, Cambridge)

\reference{} Garnavich, P., \& Szkody, P. 1988, \pasp, 100, 1522

\reference{} Greiner, J., \& Voges, W. 1997, private communication

\reference{} Hayes, D., \& Latham, D. W. 1975, \apj, 197, 593

\reference{} Hellier, C., Mason, K. O., Rosen, S. R., \&
Cordova, F. A. 1987, \mnras, 228, 463

\reference{} Hellier, C., Mason, K. O., \& Cropper, M. S. 1990,
\mnras, 242, 250

\reference{} Hellier, C., Cropper, M. S., \& Mason, K. O. 1991,
\mnras, 248, 233

\reference{} Horne, K., \& Marsh, T. R. 1986, \mnras, 218, 761

\reference{} Kenyon, S. J., \& Garcia, M. R. 1986, \aj, 91, 125

\reference{} Liebert, J., \& Stockman, H. S. 1985, Cataclysmic
Variables and Low-Mass X-Ray Binaries, eds.\ D.~Q. Lamb \& J. Patterson
(Reidel, Dordrecht), p.~151

\reference{} Martell, P. J., \& Kaitchuck, R. H. 1991, \apj, 366, 286

\reference{} Meinunger, L. 1976, Inf.\ Bull.\ Var.\ Stars.\ No.~1168

\reference{} Monet, D. G. 1979, \apj, 234, 275

\reference{} Mouchet, M., Bonnet-Bidaud, J.~M., Buckley, D.~A.~H., \&
Tuohy, I.~R., 1991, \aap, 250, 99

\reference{} Mukai, K. \& Corbet R. H. D. 1987, \pasp, 99, 149

\reference{} Patterson, J. 1994, \pasp, 106, 209

\reference{} Petrie, R. 1962, in Astronomical Techniques, ed.\
W.~A. Hiltner (University of Chicago Press, Chicago), p.~560

\reference{} Savage, B. D., \& Mathis, J. S. 1979, \araa, 17, 73

\reference{} Silber, A. 1986, Ph.D. thesis, MIT

\reference{} Stockman, H. S., Schmidt, G. D., Angel, J. R. P.,
Liebert, J., Tapia, S., \& Beaver, E.~A. 1977, \apj, 217, 815

\reference{} Szkody, P., 1995, Cape Workshop on Magnetic Cataclysmic
Variables, ASP Conference Series, Vol.~58, eds.\ D.~A.~H. Buckley \&
B. Warner (ASP, San Francisco), p.~54

\reference{} Warner, B. 1995, Cataclysmic Variable Stars (Cambridge 
University Press, Cambridge)

\reference{} Webbink, R. F. 1990, in Accretion-Powered Compact
Binaries, ed.\ C.~W. Mauche (Cambridge University Press, Cambridge),
p.~177

\reference{} Williams, G. 1983, \apjs, 53, 523

\end{references}
\end{document}